\documentclass[runningheads]{svmult}

\usepackage{epsfig}
\usepackage{rotating}
\usepackage{makeidx}   
\usepackage{graphicx}  
\usepackage{subeqnar}  
\usepackage{multicol}  
\usepackage{physprbb}  
\makeindex             

\begin{document}
\title*{Time-of-Flight Spectroscopy of Muonic Hydrogen
Atoms and Molecules%
\footnote{Contribution to {\em Hydrogen Atom II --- Precise
Physics with Simple Atomic Systems} (Springer, Berlin).} }
\toctitle{Time-of-Flight Spectroscopy of Muonic Hydrogen Atoms and
Molecules}
%
%
\titlerunning{Time-of-Flight Spectroscopy of Muonic Hydrogen
Atoms and Molecules}
%
\author{%
M.~C.~Fujiwara\inst{1,2}
\and A.~Adamczak\inst{3}
\and J.~M.~Bailey\inst{4}
\and G.~A.~Beer\inst{5}
\and J.~L.~Beveridge\inst{6}
\and M.~P.~Faifman\inst{7}
\and T.~M.~Huber\inst{8}
\and P.~Kammel\inst{9}
\and S.~K.~Kim\inst{10}
\and P.~E.~Knowles\inst{11}
\and A.~R.~Kunselman\inst{12}
\and V.~E.~Markushin\inst{13}
\and G.~M.~Marshall\inst{6}
\and G.~R.~Mason\inst{5}
\and F.~Mulhauser\inst{11}
\and A.~Olin\inst{6}
\and C.~Petitjean\inst{13}
\and T.~A.~Porcelli\inst{14}
\and J.~Zmeskal\inst{15} (TRIUMF Muonic
Hydrogen Collaboration)}

\authorrunning{M.C. Fujiwara et al.}

\institute{%
Department of Physics and Astronomy,
 University of British Columbia,\\ Vancouver, BC, Canada V6T 2A6
\and Department of Physics, Faculty of Science, University of
Tokyo,\\ Hongo, Tokyo 113-0033 Japan
\and Institute of Nuclear
Physics, 31-342 Krakow, Poland
\and Chester Technology, Chester
CH4 7QH, England, UK
\and Department of Physics and Astronomy,
University of Victoria,\\ Victoria, BC, Canada V8W 2Y2
\and TRIUMF, Vancouver, BC, Canada, V6T 2A3
\and Russian Research Center, Kurchatov Institute, Moscow 123182,
 Russia
\and Department of Physics, Gustavus Adolphus College, St. Peter, MN
 56082, USA
\and Department of Physics and Lawrence Berkeley National
Laboratory,\\
 University of California, Berkeley, CA 94720, USA
\and Department of Physics, Jeonbuk National University, \\Jeonju
 City 560-756, S. Korea
\and Institute of Physics, University of Fribourg, CH-1700
Fribourg, Switzerland
\and Department of Physics and Astronomy,
 University of Wyoming,\\ Laramie, WY 82071-3905, USA
\and Paul Scherrer Institute, CH-5232 Villigen, Switzerland
\and Institute for Medium Energy Physics, Austrian Academy of
 Sciences,\\ A-1090 Vienna, Austria}


\maketitle              

\begin{abstract}
Studies of muonic hydrogen atoms and molecules have been performed
traditionally in bulk targets of gas, liquid or solid. At TRIUMF,
Canada's meson facility, we have developed a new type of target
system using multilayer thin films of solid hydrogen, which
provides a beam of muonic hydrogen atoms in vacuum. Using the
time-of-flight of the muonic atoms, the energy-dependent
information of muonic reactions are obtained in direct manner. We
discuss some unique measurements enabled by the new technique,
with emphasis on processes relevant to muon catalyzed fusion.
\end{abstract}

\section{Introduction}
Simple atoms such as hydrogen and helium provide effective testing
ground for fundamental theories. Studies of simple {\it exotic
atoms} could add useful information, despite considerable
challenges, both experimentally and theoretically~\cite{exotic}.
In this article, we shall focus on the reactions of muonic
hydrogen atoms and molecules, and their energy dependent
properties revealed by a newly developed time-of-flight
spectroscopy technique~\cite{marsh92,marsh93,marsh96,marsh99},
with particular emphasis on reactions related to muon catalyzed
fusion phenomena.

\section{Muon Catalyzed Fusion}

A muon is a lepton of the second generation with its mass about
207 times heavier than that of electrons, and has a finite
lifetime of 2.2 $\mu$s. It is created from the decay of a pion
which is obtained, typically, from intermediate energy ($E_k \geq
500$ MeV) collisions of a proton beam on a target nucleus.

A negative muon can participate in a variety of atomic and
molecular processes. A muonic atom is formed when a muon stops in
matter replacing an electron. A muonic atom interacting with
ordinary atoms or molecules can form a muonic molecule.
The latter in turn can result in fusion
reactions between the nuclei if the target consists of hydrogen
isotopes, a phenomenon known as muon catalyzed fusion
($\mu$CF)~\cite{review}.

A single muon stopped in a target of deuterium-tritium mixture can
catalyze more than 100 fusions, but this number is limited by two
major bottle-necks. One is the rate at which a muon can go through
the catalysis cycle before its decay (cycling rate), and another
is a poisoning process called $\mu$--$\alpha$ sticking in which,
with a probability $\omega _s \leq 0.01$, the muon gets captured
after the fusion reaction to atomic bound states of the fusion
product $^4$He, and hence lost from the cycle (see
Section~\ref{sec:sticking}).

The first bottle-neck, cycling rate is determined mainly by the
rate of formation of muonic molecule $d\mu t$. In fact, a
straightforward mechanism for molecular formation via Auger
process:
\begin{equation}
\label{eq:Auger}
  \mu t + \mathrm{DX} \rightarrow [(d\mu t)xe]^+ + e
\end{equation}
is much too slow, yielding the fusion efficiency of the
order of only one fusion per muon. Here, DX refers to either
D$_2$, DT or HD molecule and $x$ the nucleus $p$, $d$ or $t$.
A resonant mechanism, however, can give much higher rates when
certain conditions are
satisfied. In the resonant formation:
\begin{equation}\label{eq:res}
\mu t + \mathrm{DX} \rightarrow [(d\mu t)_{11}xee]_{\nu K},
\end{equation}
the reaction product is a hydrogen-like molecular complex $[(d\mu
t)xee]$ in which $(d\mu t)^+$ plays a role of one of the nuclei.
The process is resonant, because the energy released upon
formation of $d\mu t$ plus the $\mu t$ kinetic energy must match
the ro-vibrational ($\nu K$) excitation energy of $[(d\mu t)xee]$
in the final state. This is only possible due to the existence of
a state $(J,v)=(1,1)$ in $d\mu t$ which is bound very loosely (in
the muonic scale) with the binding energy smaller than the
dissociation energy of deuterium molecules.

\begin{figure}[t]
  \begin{center}
    \leavevmode
 \begin{sideways}
    \epsfig{file=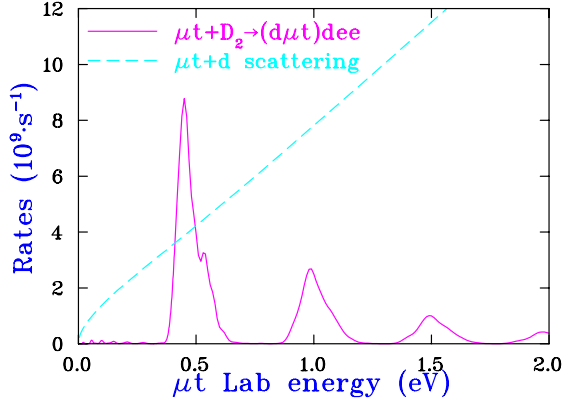,height=0.6\textwidth}
 \end{sideways}
    \caption[Resonant molecular formation rate at epithermal collision
    energies.]{Resonant molecular formation rate in $\mu t + \mathrm{D_2}$
      collisions calculated for a 3 K target~\cite{faifm89,faifm91,faifm96}.
      The rates are normalized to the liquid hydrogen density
      and averaged over $\mu t$ hyperfine states. Also shown is the
      $\mu t$ elastic scattering rate on the $d$ nucleus~\cite{chicc92}.}
  \label{fig:mut_d2}
  \end{center}
\end{figure}

Theoretical
calculations~\cite{faifm89,faifm91,faifm96,petrov94,petrov96}
predict strong enhancement of the formation rate $\lambda _{d\mu
t}$ at $\mu t$ energy of order 1 eV (Fig~\ref{fig:mut_d2}), but
direct experimental information on such epithermal resonance is
scarce, let alone its detailed structure.
Traditionally, $\mu$CF experiments in general, and molecular
formation rate measurements in particular, have been performed in
a bulk target made of mixture of hydrogen isotopes, in which
complex chains of muon induced reactions take place, making the
interpretation difficult and model-dependent.
In addition, the resonant energies
$\sim$1 eV is difficult to access with the target thermal energy,
since it would require a target of several thousand degrees, a
formidable task when working with tritium.

At TRIUMF we have developed a new target system which uses
multilayer of solid hydrogen thin films~\cite{knowl96,knowl93}.
The use of $\mu t$ beam obtained from the thin film target
provides some unique advantages in tackling the epithermal resonances
experimentally: (1) formation process can be isolated, to large
extent, from the rest of the cycle, (2) epithermal energies are
directly accessible due to the available beam energy, and (3) $\mu
t$ time of flight across the drift distance provide information of
the resonance energies. Obviously, many technical challenges had
to be overcome to make use of the new method.

\section{A beam of muonic hydrogen atoms}

\begin{figure}[t]
  \begin{center}
    \leavevmode
    \epsfig{file=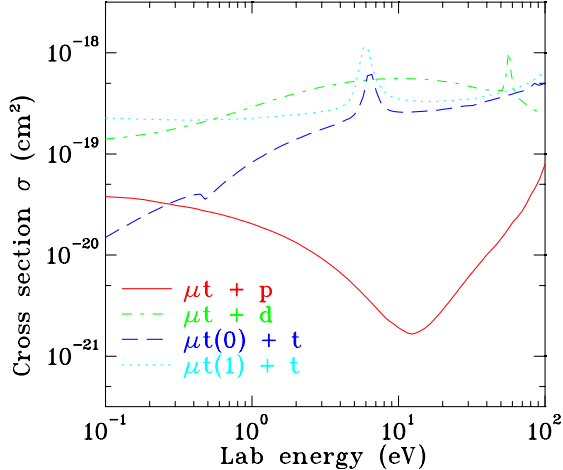,width=0.6\textwidth}
    \caption[Scattering cross sections of muonic tritium with a hydrogen
    isotope nucleus.]{Scattering cross sections for $\mu t$ with a hydrogen
      isotope nuclei from Refs.~\cite{chicc92,bracc89}, showing
      the Ramsauer-Townsend minimum at around 10 eV for $\mu t + p$. $\mu t
      (F) + t$ cross sections plotted include both elastic and spin exchange
      reactions, where $\mu t$(0) is the singlet state and $\mu t$(1) is
      the triplet state.}
  \label{fig:RT_atlas}
  \end{center}
\end{figure}

The basic processes involved in creating a beam of muonic tritium
atoms~\cite{fujiw97} can be categorized into four
step~\cite{marku96}: {\em atomic formation}, {\em acceleration},
{\em extraction}, and {\em moderation}. When a muon is stopped in
a thin solid hydrogen target consisting of protium ($^1$H$_2$)
doped with a small amount ($c_t\sim 0.1$\%) of tritium, muonic
protium atom ($\mu p$) is mostly formed ({\em atomic formation}).
The muon quickly transfers from proton to triton~\cite{mulha96} to
form $\mu t$, the latter more tightly bound due to the reduced
mass difference. In the reaction, the $\mu t$ gains about 45 eV of
recoil kinetic energy, and is thus {\em accelerated}. The $\mu t$
then slow down from the collisions with the rest of the target
nuclei (mostly protons), until it reaches about 10 eV. At these
energies, $\mu t + p$ elastic scattering cross section drops
dramatically due to the Ramsauer-Townsend effect, making the rest
of the target nearly transparent (Fig~\ref{fig:RT_atlas}). The
$\mu t$ atom is thus {\em extracted} from the layer into vacuum.
The energy of the emitted $\mu t$, which is close to the
Ramsauer-Townsend minimum of $\sim$10 eV~\cite{fujiw99}, can be
controlled to some extent by placing an additional layer, for
example of deuterium, on top of the emission layer ({\em
moderation}). Creation of muonic deuterium ($\mu d$) is possible
in a similar manner with a deuterium-doped protium target: in
fact, the emission of muonic hydrogen atoms was first discovered
in this system~\cite{forst90}. See
Refs.~\cite{jacot96,knowl97,olin99} for our measurements involving
muonic deuterium. Recently, we also observed emission of muonic
protium atoms from pure hydrogen layer~\cite{wozni99}, though
emission mechanism is expected to be completely
different~\cite{adamc99}.


\section{Resonant formation of muonic molecules $d\mu t$}

\begin{figure}[t]
 \begin{center}
 \epsfig{file=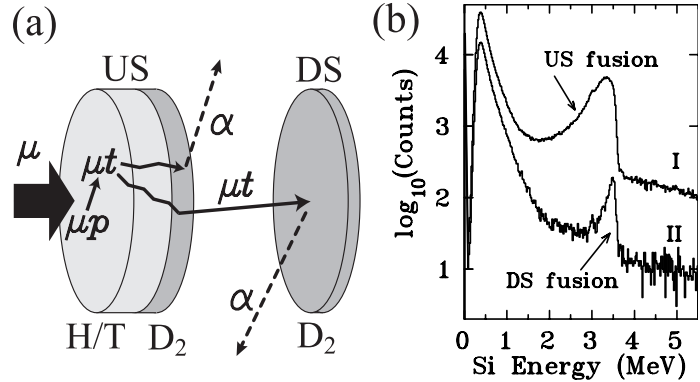,width=0.75\textwidth}
  \end{center}
\caption{(a) Schematic of the thin film target for the $d\mu t$
formation measurements consisting of the emission, the moderation
and the reaction layers, which are prepared by rapidly freezing
hydrogen isotopes on the gold foils (not shown) held in vacuum at
3.5 K~\protect\cite{knowl96}. The layer thickness (3.43
mg$\cdot$cm$^{-2}$, 96 $\mu$g$\cdot$cm$^{-2}$, and 21
$\mu$g$\cdot$cm$^{-2}$, respectively) were measured off-line via
$\alpha$ particle energy loss (see Fig.~\ref{fig:thickness}) (b)
Measured Si energy spectra with prompt (I: $t>0.02$ $\mu$s) and
delayed (II: $t>1.5$ $\mu$s) time cuts. Fusion in DS reaction
layer is separated from that in US D$_2$ due to the $\mu t$ TOF
across the vacuum.}
 \label{fig:siE}
\end{figure}

Figure~\ref{fig:siE} (a) illustrates the principle of our
molecular formation measurements using the atomic beam
method~\cite{marsh92,marsh93,marsh96,marsh99}. A beam of $5\times
10^3$ $\mu ^-$/s of momentum 27 MeV/$c$ from the TRIUMF M20B
channel was degraded in a 51 $\mu$m gold target support, and
stopped in the upstream (US) emission layer. The $\mu t$ beam,
obtained as described above, were slowed via elastic scattering in
a D$_2$ moderation layer from some 10 eV to near 1 eV to better
match the resonance energies. The $\mu t$, after a few $\mu$s time
of flight (TOF), is collided with a reaction layer in downstream
(DS), separated by the drift distance of 17.9 mm in vacuum.
Formation of $d\mu t$ molecules is detected by observing 3.5 MeV
$\alpha$ particles produced in the fusion reaction, $d + t
\rightarrow \alpha + n$, which follows the formation. Si detectors
placed in the vacuum enables the measurement $\alpha$ with high
energy resolution. The background can be determined accurately by
``turning off'' the DS fusion reactions using the target without
the DS layer. This ability to control a  specific process, without
affecting the rest, is an advantage of the thin film method. (In
conventional targets, changing the target conditions would affect
many processes simultaneously).

\begin{figure}[t]
  \begin{center}
    \leavevmode
 \begin{sideways}
    \epsfig{file=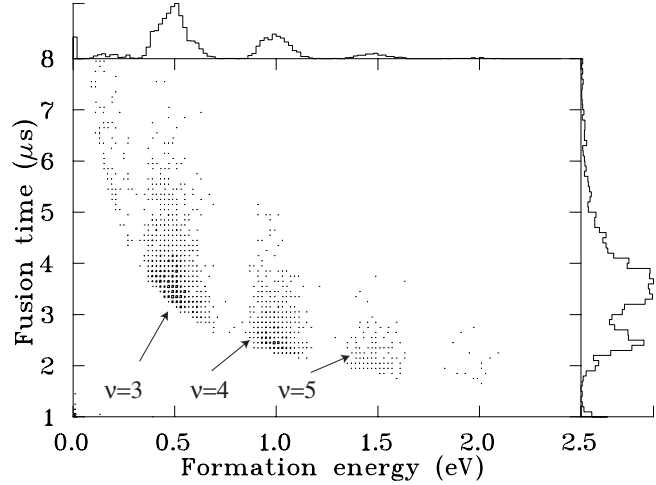,height=0.7\textwidth}
 \end{sideways}
    \caption[Resonant molecular formation rate at epithermal collision
    energies.]{Simulated correlations between the fusion time and the
    energy at which molecular formation takes place when there is no
    $\mu t$ energy loss in DS layer. Corresponding vibrational excitation
    in the [$(d\mu t)dee$] system are also shown.}
  \label{fig:tofscat}
  \end{center}
\end{figure}

Because the time between muon stop and the fusion $\alpha$
detection (which we call {\em fusion time}) is dominated by the
$\mu t$ TOF, it provides a measure on molecular formation energy,
as long as the energy loss of $\mu t$, due to elastic scattering
before the formation in the DS D$_2$, is small.
Figure~\ref{fig:tofscat} illustrates the simulated correlation
between the fusion time and the energy at which molecular
formation takes place, when there is little $\mu t$ energy loss. A
thin DS layer ($\sim$1 $\mu$m)
was thus used in the measurement to minimize the energy
loss and maximize the time-energy correlation.

\begin{table}[b]
  \begin{center}
    \leavevmode
    \caption{Estimated effects on the formation rate scaling parameter
     ${\cal S}_\lambda$ by the systematic uncertainties
     in the MC modelling.}
    \begin{tabular}{lr} \hline \hline
MC error source & $\Delta {\cal S}_\lambda/{\cal S}_\lambda$ \%
 \\ \hline
 $\mu$ beam size & 1.2 \\
 Nonuniform $\mu$ stopping (GEANT) & 1.8 \\
 $\mu t$ TOF drift distance & 2.6 \\
 $\mu p \rightarrow \mu t$ transfer, $p\mu p$ formation & 5.6 \\
 $\mu t + d$, $\mu t + t$ scattering, layer thickness & 5.7 \\
 $\mu t + p$ RT minimum energy  & 1.1 \\
 Resonance Doppler widths in solid & 8.0 \\
 Solid state and low energy processes & 5.0 \\
 (subthreshold resonances, slow thermalization, & \\
 $\mu t$ energy after resonant scattering) & \\ \hline
 Total (in quadrature)  &  12.9 \\ \hline \hline
    \end{tabular}
    \label{tab1}
  \end{center}
\end{table}

\begin{figure}[h]
 \begin{center}
 \fbox{\epsfig{file=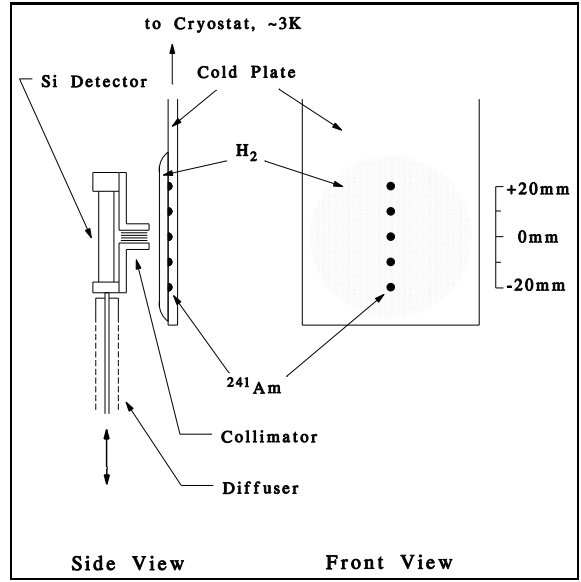,width=0.6\textwidth}}
 \end{center}
 \caption{Calibration measurement of solid hydrogen film layer
 thickness via energy loss of $\alpha$
 particles~\protect\cite{fujiw97b,fujiw96b}. Americium $\alpha$ source is
 embedded on the surface of the gold-plated copper substrate,
 onto which hydrogen thin film is deposited by releasing the gas
 through porous sintered metal (diffuser). Silicon detector,
 mounted on the vertically movable diffuser, measured the
 $\alpha$ particle energy loss in the film, which is converted the
 thickness using the stopping power.}
 \label{fig:thickness}
 \end{figure}

\begin{figure}[h]
 \begin{center}
 \epsfig{file=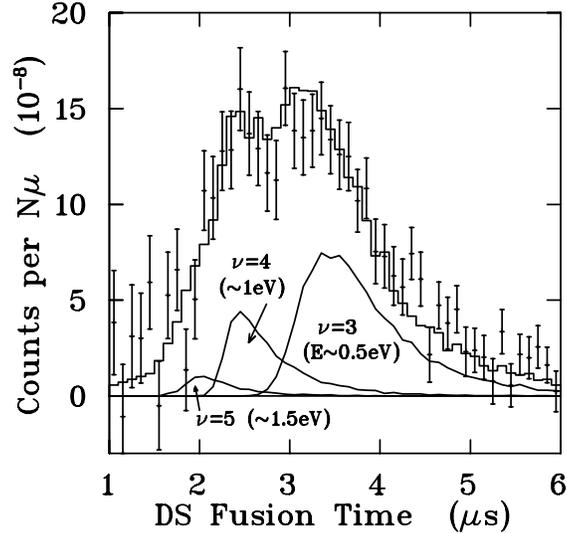,width=0.6\textwidth}
 \end{center}
 \caption{Time-of-flight fusion spectrum (error bars) and
 simulation spectrum (histogram), normalized the number of
 incident muons $N_\mu$. Also plotted are simulated
 contributions from different resonance peaks given by
 the time--energy correlated events.}
 \label{fig:mc}
 \end{figure}

A great deal of efforts were made to understand the systematics of
this new method. Fig.~\ref{fig:thickness} and Table~\ref{tab1}
give some such examples. The detail of the analysis can be found
in Refs.~\cite{phd,prl}. The resulting DS fusion time spectrum and
its comparison with Monte Carlo (MC) simulations are shown in
Fig.~\ref{fig:mc}, which clearly establishes the resonance
structure. From the time-of-flight analysis of $2036\pm 116$ DS
fusion events, a formation rate consistent with $0.73\pm
(0.16)_{meas} \pm (0.09)_{model}$ times the theoretical prediction
of Faifman {\it et al.}~\cite{faifm96} was obtained (the first
error refers to the measurement uncertainty including the
statistics and the second is that in MC modeling). The resonance
energies were determined from the fit to be $0.940\pm
(0.036)_{meas} \pm (0.080)_{model}$ times the
theory~\cite{faifm96}. Thus, for the first time, the existence of
epithermal resonances in $d\mu t$ molecular formation was directly
confirmed, and their energies measured. For the largest peak at
the resonance energy of $0.423\pm 0.037$ eV, our results
correspond to the peak rate of $(7.1 \pm1.8)\times 10^9$ s$^{-1}$.
This is more than an order of magnitude larger than the rates at
lower energies, experimentally demonstrating the prospect for high
cycling $\mu$CF in a high temperature target of several thousand
degrees. If one assumes the energy levels of the [$(d\mu t)dee$]
molecule, which have a similar structure to those of ordinary
hydrogen, our results imply sensitivity to the binding of energy
of the loosely bound $(d\mu t)_{11}$ state with an accuracy
comparable to the vacuum polarization and other QED corrections,
opening up a new possibility of precision spectroscopy in a
quantum few body system.

Recently $\mu$CF using a triple isotope mixture (H/D/T) target has
drawn considerable interest~\cite{triple}, particularly because of
resonances in the $d\mu t$ formation in the $\mu t + \mathrm{HD}$
collisions, predicted to be even stronger than the $\mu t +
\mathrm{D_2}$ case. Our thin film target allows us to create $\mu
t$ collision with pure HD molecules, and time-of-flight method
described here should give a direct test of the theoretical
prediction.
The data for the $\mu t + \mathrm{HD}$ collision have been
collected, and the result will be reported in our future
publication~\cite{porce00}.

\section{$\mu$--$\alpha$ sticking: yet another bottle-neck}
\label{sec:sticking}

Although our measurements of epithermal resonant formation
indicates molecular formation may no longer be a bottle-neck at
appropraite conditions, the process of $\mu$-$\alpha$ sticking
still gives a stringent limit, independent of the cycling rate, on
the number of fusions catalyzed by one muon. Intense efforts have
been made for nearly two decades to understand this key process,
but discrepancies persist between and theory and experiment
(including latest PSI~\cite{petit93} and RIKEN-RAL~\cite{riken}
results), the experiment being systematically lower than the
theory~\cite{stick-th0,stick-th}.

 \begin{figure}[t]
 \begin{center}
 \epsfig{file=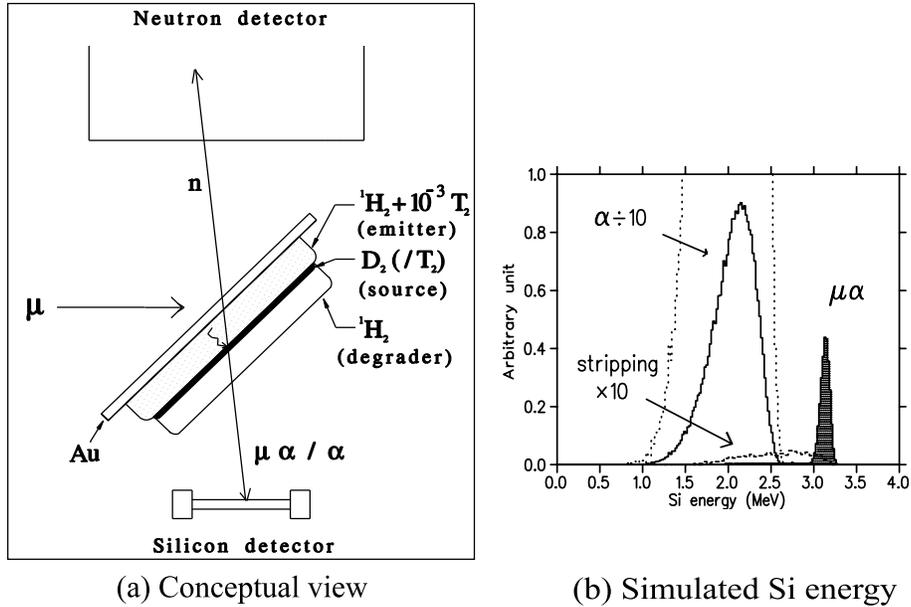,width=\textwidth}
 \end{center}
 \caption{Proposed direct measurement of $\mu$-$\alpha$
 sticking~\protect\cite{fujiw96,eec}.
 (a) The $\mu t$ created in the {\it emitter} is stopped
 in the {\it source}, where fusion takes place producing $\alpha ^{++}$
 or ($\mu \alpha$)$^+$. The {\it degrader} separates the two species
 by their stopping powers (recall that $dE/dx \propto
 Z^2$). Collinear coincidence of the neutron with the charged events
 suppresses the background. (b) Simulated Si
 spectrum for the coincidence events, showing a clear separation
 between $\alpha$ and $\mu \alpha$ peaks.}
 \label{fig:stick}
 \end{figure}

Theoretical calculations of sticking are challenging, due to the
interplay of the Coulomb and strong interactions in a
non-adiabatic few-body system, yet recent predictions, including
the effects of nuclear structure and the deviations from the
standard sudden approximation, now converge to a few
percent~\cite{stick-th}.
They cannot, however, be readily compared to
experiment because most measurements are primarily sensitive to
{\it final sticking} ($\omega _s^{fin}$), which is a combination
of {\it initial sticking} ($\omega _s^{0}$), the intrinsic
branching ratio for $d\mu t \rightarrow \mu \alpha +n$, and {\it
stripping} ($R$), collisional reactivation of the muon from $\mu
\alpha$ in the target medium ({\it i.e.}, $\omega _s^{fin} \equiv
\omega _s^{0} (1-R)$).

Discrepancies in the experimental sticking results has been in
part attribu\-ted~\cite{petit93} to the difficulties associated
with the conventional neutron method, which include a high model
dependence and the need for the absolute neutron calibration. A
recent RIKEN-RAL experiment~\cite{riken} has directly observed,
with impressive statistics, X-rays from $\mu \alpha$ deexcitaion
following the muon sticking to excited states of $\mu \alpha$,
taking advantage of the new intense pulsed muon beam which enabled
the large signal enhancement, but unfortunately the determination
of the sticking probability has to rely on the models of $\mu
\alpha$ cascade and stripping, which are yet largely untested.

We have proposed a new direct experiment of sticking using our
multi-layer thin film target. The method is illustrated in
Fig~\ref{fig:stick}, and the details are given in
Refs.~\cite{fujiw96,eec}. The determination of sticking from the
ratio $\mu \alpha / (\mu \alpha + \alpha)$ is simple and model
independent. Since the stripping in the degrader is small, the
measurement is sensitive to the initial sticking, while the
stripping itself process can be systematically studied by varying
the degrader thickness. Thus experimental separation of initial
sticking and stripping will become possible for the first time.

\section{Future prospects}

Currently there is growing interest in the high energy physics
community world-wide in developing a muon storage ring for a
neutrino factory and/or for a muon collider~\cite{himus99}. Low
energy muon science, including muonic atom and molecule studies
would greatly benefit, if these facilities are realized. Some of
the experiments with solid thin films targets are presently
limited by the statistics, due to small fraction of muon stops in
the target layer, and increased background from muon stopping in
the target support and the vacuum chamber~\cite{fujiw00b}. We can
expect substantial improvements in the accuracy, as well as
entirely new types of experiments with the advent of high
intensity muon sources.

\section*{Acknowledgements}

This work is supported in part by Canada's NSERC and NRC, USA's
DOE and NSF, Swiss National Science Foundation, and NATO Linkage
Grant. M.C.F. thanks the support of Rotary Foundation, UBC, Green
College, Government of Canada, Westcoast Energy, Nortel and Japan
Society of the Promotion of Science.

%

\end{document}